\documentclass[12pt]{article}
\usepackage{graphicx}
\usepackage[russian,english]{babel}
\usepackage[LCY]{fontenc}
\usepackage{amssymb}
\usepackage{amsmath}
\usepackage{fullpage}
\usepackage[cp1251]{inputenc}

\begin{document}

\begin{center}
\Large Standard Model in adS slice with UV-localized Higgs field
\end{center}

\begin{center}
Sergey Mironov$^{a,b}$, Mikhail Osipov$^{c,d}$, Sabir
Ramazanov$^{a,c}$

\emph {$^{a}$Moscow State University, Department of Physics, \\
Vorobjevy Gory, 119991, Moscow, Russia\\$^{b}$Institute of
Theoretical and Experimental Physics, \\B. Cheremushkinskaya ul.
25, 117259 Moscow, Russia
\\$^{c}$Institute for Nuclear Research of the Russian Academy of Sciences,\\
60th October Anniversary prospect 7a, Moscow, 117312, Russia\\$^{d}$Moscow Institute of Physics and Technology,\\
Institutskii per., 9, Dolgoprudny, 141700 Moscow Region, Russia }

\end{center}
\begin{center}
\Large Abstract
\end{center}
We discuss five-dimensional Standard Model in a slice of adS
space-time with the Higgs field residing near or on the UV brane.
Allowing fermion fields to propagate in the bulk, we obtain the
hierarchy of their masses and quark mixings without introducing
large or small Yukawa couplings. However, the interaction of
fermions with the Higgs and gauge boson KK excitations gives rise
to FCNC with no built-in suppression mechanism. This strongly
constrains the scale of KK masses. We also discuss neutrino mass
generation via KK excitations of the Higgs field. We find that
this mechanism is subdominant in the scenarios of spontaneous
symmetry breaking we consider.

\section{Introduction and summary}
Five-dimensional theories with the Standard Model fields living in
adS slice attract considerable interest, especially due to their
possible connection to adS/CFT correspondence \cite{Maldacena,
Gubser, Witten}. Originally, theories of this sort focused on the
problem of the hierarchy between the Planck and electroweak scales
\cite{Sundrum}. Later on, it has been understood that they can
explain the hierarchy of fermion masses and quark mixings without
introducing large or small parameters into the original action
\cite{Gherghetta, Huber, Stephan, Grossman, Chang}.

AdS slice is a solution to the Einstein equations in 5-dimensional
(5D) space-time with two gravitating branes. Upon fine-tuning the
brane tension and 5D cosmological constant, one obtains the metric
\begin{equation}
ds^2=e^{-2k|y|}\eta_{\mu \nu}dx^{\mu}dx^{\nu}-dy^2 \; ,
\label{metrics}
\end{equation}
where $k$ is the adS curvature, $y$ denotes the coordinate of the
fifth warped dimension, which is $S^1/Z_2$ orbifold of size $R$.
Two branes are placed at $y=0$ and $y=\pi R$. These are
ultraviolet (UV, $y=0$) and infrared (IR, $y=\pi R$) branes,
respectively. We choose the 4-dimensional (4D) Minkowski metric as
$\eta_{\mu \nu}=(1,-1,-1,-1)$. This setup is known as the
Randall-Sundum model of type one (RS1)  \cite{Sundrum}. In the
original RS1 model, only gravity is supposed to propagate in the
5D space-time, while the SM fields reside on the IR brane.
However, this is not the only possibility, as one can allow all
particles or some of them propagate in the bulk. In that case, the
5D fields can be expanded in the tower of the Kaluza--Klein (KK)
modes, and the zero modes are associated with the SM fields. The
hierarchy of fermion masses and quark mixings is then obtained by
an appropriate choice of the fermion localization in the bulk.

It is most common to assume that the Higgs field is localized on
or near the IR brane, for reviews see, e.g., Refs.
\cite{Casagrande, Neubert}. In that case, the zero modes of light
fermions are localized near the UV brane, while zero modes of
heavy quarks are localized towards the IR brane. The smallness of
the SM Yukawa couplings is then due to the small overlaps of the
zero modes of the light fermions and the Higgs field. Masses of KK
modes of the gauge and Higgs fields (if the latter lives in the
bulk) are constrained by the requirement of FCNC suppression. In
models with the Higgs field localized on or near the IR brane,
this constraint is fairly weak: excited KK modes can have masses
of order 10 TeV. This is due to the so called RS--GIM mechanism
\cite{Perez, Agashe}, which is built-in: since the zero modes of
light fermions are localized near the UV brane, their overlaps
with the KK modes of the gauge and Higgs fields are exponentially
small. Moreover, introducing additional symmetries, it is possible
to relax the constraints on KK masses down to about $3$ TeV
\cite{Csaki, Blanke, Randall, Santiago}. Thus, effects of new
physics can be observed in experiments at LHC.

According to the holographic picture \cite{Contino, lesh, Arkani},
every bulk zero mode field corresponds to an  eigenstate in the
dual 4D theory, which is a mixture of elementary source and
composite CFT fields. If the bulk zero mode is localized towards
the UV brane, the massless eigenstate of the dual theory is
predominantly the source field. Conversely, the dual
interpretation of a bulk zero mode localized towards the IR brane
is a state that is predominantly a CFT bound state. If the Higgs
field is confined to the IR brane, it is interpreted as a pure CFT
bound state in the dual theory. The top-quark is mostly a CFT
bound state, while light fermions are mostly elementary.

In this paper we turn this picture upside down and consider a
scenario with the bulk Higgs field localized near or on the UV
brane. Without introducing the hierarchy in the parameters of the
original action, we show that the realistic pattern of fermion
masses and quark mixings can be obtained in this case as well.
However, the overall picture is quite different. Namely, light
quarks and right leptons are localized near the IR brane to have
small overlaps with the UV-localized Higgs field and,
consequently, small Yukawa couplings. Hence, many light SM fields
are mostly CFT composites in the dual picture, while heavy fields
are predominantly elementary.

The IR localization of light fermions introduces, however, the
FCNC problem. Indeed, the KK excitations of the Higgs field and
bulk gauge bosons also live near the IR brane. Thus, their wave
functions have large overlaps with the wave functions of light
fermions. From the point of view of the effective 4D theory, this
means that the corresponding couplings are of order one. Then the
only parameter one can use for suppressing FCNC is the mass scale
of KK excitations. The latter must be very high to satisfy the
existing constraints coming from kaon mixing. We will see that the
constraint on the KK scale is particularly strong for the Higgs
field living in the bulk.

A way to avoid FCNC mediated by the Higgs KK excitations is to
localize the Higgs field on the UV brane. Then there are no Higgs
KK excitations at all. In that case the dominant source of FCNC is
the exchange by the KK excitations of the gauge fields. Although
the constraints here are less severe, the allowed scale of the KK
excitations is still beyond the experimental reach.

We also discuss neutrino masses of the Dirac type. It is
straightforward to obtain them via the interaction with the zero
mode of the Higgs field. An alternative possibility, which would
probably be more interesting, would be the neutrino mass
generation via the interaction with the KK excitations of the
Higgs field. If it worked, the smallness of the neutrino masses
would be due to the suppression of the vacuum expectation values
of the heavy KK Higgs modes, rather than due to small effective 4D
Yukawa couplings. In the particular scenarios we consider, this
mechanism is subdominant, however: the neutrino interactions with
the Higgs zero mode are always strong enough to generate the main
contribution to the neutrino masses.

This paper is organized as follows. In Section 2 we discuss
possible scenarios of spontaneous symmetry breaking in the 5D
Standard Model with the Higgs field localized towards the UV
brane. In Section 3 we show that realistic 4D fermion masses and
quark mixings can be obtained without introducing small or large
parameters in the 5D action. There we also discuss neutrino masses
of the Dirac type. We consider the FCNC problem in Section 4. We
find that the constraints on the mass scale is $m_{KK}\gtrsim
5\times 10^{5}$ TeV for the Higgs field living in the bulk and
$m_{KK} \gtrsim 700$ TeV for the Higgs field localized on the UV
brane. This reiterates the power of the FCNC constraints in models
without a built-in mechanism of the FCNC suppression.

\section{Scenarios of electroweak symmetry breaking}
We consider the Standard Model in the slice $y\in (0,\pi R)$ of 5D
adS space-time with the metric \eqref{metrics}. We will see in
Section 3 that the fermion mass hierarchy problem is naturally
solved provided that
\begin{equation}\label{large}
e^{k\pi R}\gg 1\; .
\end{equation}
We treat $e^{k\pi R}$ as a large parameter in what follows.

The action for the Higgs field living in the 5D bulk is
\begin{equation} \label{spsym1}
S_{5}= \int d^{4}x dy \sqrt{g}\Bigl(\frac{1}{2}g^{M
N}{\partial}_{M} H {\partial}_N H -\frac{1}{2}m^2_{H}
{H}^2-V(H)\Bigr)+S_{b}\; ,
\end{equation}
where $m_H$ is the bulk Higgs mass, $V(H)$ is the symmetry
breaking potential. The brane term $S_{b}$ is added to have the
zero mode in the absence of the potential $V(H)$
\cite{Gherghetta},
\begin{equation}\label{braneterm}
\nonumber S_{b}=\Bigl(1-\frac{\alpha}{2}\Bigr)k\int
d^4xdy\sqrt{g}(\delta (y-\pi R)-\delta(y)){H}^2\; .
\end{equation}
The constant $\alpha$ is tuned to $\alpha
=\sqrt{4+\frac{m^2_H}{k^2}}$, so that the zero mode exists. We
will momentarily see that with the negative sign in front of
$\alpha$ chosen in \eqref{braneterm} and $\alpha >1$, the Higgs
zero mode is localized near the UV brane. This is the case we
study in what follows.

Let us first switch off the potential $V(H)$, i.e., set $V(H)=0$,
and consider the free scalar field. One derives from the 5D action
\eqref{spsym1} with the brane term \eqref{braneterm} the following
equations of motion and boundary conditions,
\begin{equation}
\partial_{\mu} \partial^{\mu} H+e^{2ky}\partial_5(e^{-4ky}\partial_5 H)+m^2_He^{-2ky}H=0\; ,
\end{equation}
\begin{equation}\label{bounds}
\partial_5 H-(2-\alpha)k H|_{0,~\pi R}=0\; .
\end{equation}
Following the standard procedure, we expand the field $H$ in the
infinite sum:
\begin{equation}\label{scKK}
H (x,y)=\sum_{n=0}^{\infty} {h}_{n} (x) {H}_{n} (y)\; ,
\end{equation}
where $h_{n}(x)$ are KK modes with masses $m_{n}$, while
$H_{n}(y)$ are their bulk profiles. The zero mode is given by
\cite{Gherghetta}
\begin{equation}\label{scprofiles}
H_{0}(y)=N_{0}e^{(2-\alpha) ky}\; .
\end{equation}
It is clear from \eqref{spsym1} that the effective profile is
actually $e^{-ky}H_{0}(y)$. Hence, the zero mode is UV-localized
for $\alpha >1$. The normalization constant $N_{0}$ ensures the
standard form of the kinetic term in the effective 4D action. The
latter condition reads
\begin{equation}\label{norm}
\int^{\pi R}_0 dye^{-2ky}H_{n}^2(y)=1\; ,
\end{equation}
so that
\begin{equation}\label{scnorm}
N_{0}=\sqrt {\frac{2k(\alpha-1)}{1-e^{2(1-\alpha)k\pi R}}} \approx
\sqrt{2k(\alpha -1)}\; .
\end{equation}

The profiles of the excited KK modes are given by
\cite{Gherghetta}
\begin{equation}\label{scprofn}
H_{n}(y)=N_{n}
e^{2ky}\left[J_{\alpha}\Bigl(\frac{m_{n}}{k}e^{ky}\Bigr)+
\frac{J_{\alpha-1}(\frac{m_{n}}{k})}{J_{-\alpha+1}(\frac{m_{n}}{k})}J_{-\alpha}
\Bigl(\frac{m_{n}}{k}e^{ky}\Bigr)\right]\; ,
\end{equation}
with the normalization constants
\begin{equation}\label{scnormn}
N_{n}\approx \frac{m_{n}}{\sqrt{k}}
\frac{1}{\sqrt{{\int}^{\beta_n}_0 sJ^2_{\alpha}(s)ds}}\; .
\end{equation}
Here
\begin{equation}
\beta_n=\frac{m_n}{k}e^{k\pi R}\; .
\end{equation}
The boundary conditions \eqref{bounds} determine the eigenvalues
$\beta_n$ and hence the masses of the KK excitations; these are
found from
\begin{equation}
J_{\alpha-1}(\beta_n)=0\; .
\end{equation}
Clearly, the lowest KK modes have $\beta_n\sim 1$ and hence
\begin{equation}
m_n\sim ke^{-k\pi R}\; .
\end{equation}
Note that $\frac{m_n}{k}$ is a small parameter for not too large
values of $n$ in the regime \eqref{large} we consider.

To obtain the Higgs VEV, we turn on the potential $V(H)$. Let us
begin with the choice
\begin{equation}
V(H)=-\frac{{\mu}^2}{2} {H}^2+\lambda {H}^4 \; ,
\end{equation}
so that symmetry breaking occurs due to the bulk mass term. By
inserting the KK decomposition \eqref{scKK} into the action
\eqref{spsym1} and integrating over the fifth coordinate, one
obtains the effective 4D action. Assuming that the KK excitations
are small, we treat the interaction between the  zero modes and
excited KK modes in the linear approximation in $h_{n}$ and write
\begin{equation}\label{spsymeff1}
\begin{split}
 S_{eff}&=\int d^4 x \Bigl(\frac{1}{2}({\partial h_0})^2
+\sum_{n=1}^{\infty} \frac{1}{2}({\partial {h}_n})^2
-\sum_{n=1}^{\infty} \frac{1}{2}m^2_{n}{h}^2_n+
\frac{1}{2} {\mu}^2(c_0 h_0^2+ \sum_{n=1}^{\infty} 2 c_n {h}_0{h}_n)\\
&\quad -\lambda (a_0 {h}^4_0+\sum_{n=1}^{\infty} 4a_n {h}^3_0
{h}_n)\Bigr)\; ,
\end{split}
\end{equation}
where the constants $a_{0}$, $a_{n}$, $c_0$ and $c_n$ are the
overlap integrals
\begin{equation}\label{a0c0}
a_0=\int_{0}^{\pi R} dy \sqrt{g} {H}^4_0\; , \quad
c_0=\int_{0}^{\pi R} dy \sqrt{g}{H}^2_0\; ,
\end{equation}
\begin{equation}\label{ancn}
 a_n=\int_{0}^{\pi R} dy \sqrt{g} {H}^3_0 {H}_n \; , \quad c_n=\int_0^{\pi R} dy
\sqrt{g} {H}_0 {H}_n\; .
\end{equation}
Making  use of the effective action \eqref{spsymeff1}, we derive
VEVs of the zero and excited KK modes:
\begin{equation}\label{spsymvev0}
v_0= \sqrt{\frac{c_0{\mu}^2}{4a_0 \lambda}}\; ,
\end{equation}
\begin{equation}\label{spsymvevn}
v_n= \frac{c_n{\mu}^2v_0-4a_n\lambda v_0^3}{m^2_n}\; .
\end{equation}
As the zero mode represents the standard Higgs field, we have:
$v_0=v_{SM}= 247$ GeV. We evaluate the integrals \eqref{a0c0} and
obtain
\begin{equation}\label{barhan}
a_0\approx \frac{N^4_0}{4(\alpha -1)k}\; , \quad c_0\approx
\frac{N^2_0}{2\alpha k}\; .
\end{equation}
We see that the constants $a_0$ and $c_0$ are estimated as
$a_0\sim k$ and $c_0 \sim 1$. Then the mass parameter $\mu$
responsible for symmetry breaking is of the order of the SM Higgs
VEV, while $\lambda\lesssim k^{-1}$ in order that the effective 4D
coupling $\lambda_4=\lambda a_0$ be small.

Making use of \eqref{spsymvev0} in \eqref{spsymvevn}, we write the
KK VEVs as follows
\begin{equation}
v_{n}= (a_{0}c_{n}-a_{n}c_{0})\frac{v_0\mu^2}{a_0m^2_n}\; .
\end{equation}
In what follows, we need the integrals \eqref{ancn} to the
subleading order in $\frac{m_{n}}{k}.$ The constants $a_{n}$ are
different at $\alpha
>2$ and $\alpha <2$,
\begin{equation}\label{an}
\begin{split}
a_n&=-\frac{N^3_0 N_n}{2^{\alpha} \Gamma (\alpha)
m_n}\left(\frac{m_n}{k}\right)^{\alpha -1}\left[1-\frac{(\alpha
-1)}{2(\alpha -2)(2\alpha
-3)}\left(\frac{m_n}{k}\right)^2 \right] \qquad \alpha >2 \; ,\\
 a_n&=-\frac{N^3_0 N_n}{2^{\alpha} \Gamma (\alpha) m_n}\left(\frac{m_n}{k}\right)^{\alpha -1}\left[1-2^{\alpha}\Gamma
(\alpha)\left(\frac{m_n}{k}\right)^{2(\alpha -1)}
\int^{\beta_n}_0s^{3(1-\alpha)}J_{\alpha}(s)ds \right] \qquad
\alpha <2\; .
\end{split}
\end{equation}
The constants $c_{n}$ are given by
\begin{equation}\label{cn}
\begin{split}
c_n=-\frac{N_0 N_n}{\alpha 2^{\alpha} \Gamma (\alpha)
m_n}\left(\frac{m_n}{k}\right)^{\alpha -1}\left[2(\alpha
-1)+\left(\frac{m_n}{k}\right)^2 \ln \frac{m_n}{k}\right]\; .
\end{split}
\end{equation}
The constants $a_{n}$ and $c_{n}$ are estimated as $a_{n}\sim
k\left(\frac{m_n}{k}\right)^{\alpha -1}$ and $c_{n}\sim
\left(\frac{m_n}{k}\right)^{\alpha -1}$. Then a naive estimate of
the KK VEVs would be $v_{n}\sim
\left(\frac{v_{SM}}{m_n}\right)^2\left(\frac{m_n}{k}\right)^{\alpha
-1}v_{SM}$. However, this is not the case. Indeed, to the leading
order in $\frac{m_n}{k}$ the constants satisfy
\begin{equation}
\frac{a_0}{c_0}=\frac{a_n}{c_n}\; .
\end{equation}
Taking into account the subleading terms in Eqs.~\eqref{an},
\eqref{cn}, we obtain the following expressions\footnote{One can
show that the terms omitted in \eqref{barhan} are negligible.} for
the KK VEVs at $\alpha
>2$ and $\alpha <2$:
\begin{align}\label{vevnn}
v_n&= \frac{1}{2^{\alpha} \Gamma{(\alpha +1)}}
\sqrt{\frac{2(\alpha -1)}{\int^{\beta_n}_0 sJ^2_{\alpha}(s)ds}}
\left(\frac{\mu}{m_n}\right)^2
\left(\frac{m_{n}}{k}\right)^{\alpha +1}\ln
\frac{k}{m_n}v_0 \qquad \alpha >2 \; ,\\
v_n&=  \sqrt{\frac{8(\alpha -1)^3}{\int^{\beta_n}_0
sJ^2_{\alpha}(s)ds}} \left(\int^{\beta_{n}}_{0}s^{3(1-\alpha)}
J_{\alpha}(s) ds\right)\left(\frac{m_{n}}{k}\right)^{3(\alpha
-1)}\frac{{\mu}^2}{\alpha m^2_n}v_0 \qquad \alpha <2\; .
\end{align}
So, modulo factors of order one, the estimates are $v_{n}\sim
\left(\frac{v_{SM}}{m_n}\right)^2
\left(\frac{m_{n}}{k}\right)^{\alpha +1}v_{SM}$ in the case of the
Higgs field localized with the parameter $\alpha>2$ and
$v_n\sim\left(\frac{v_{SM}}{m_n}\right)^2\left(\frac{m_{n}}{k}\right)^{3(\alpha
-1)}v_{SM}$ for $\alpha <2$.

So strong suppression is absent in a model with another mechanism
of spontaneous symmetry breaking. Instead of the potential
\eqref{spsym1}, one introduces
\begin{equation}\label{spsym2}
V(H)=-\frac{1}{2}M\delta (y)H^2+\lambda H^4\; ,
\end{equation}
so that the mass term resides on the UV brane. Proceeding as
before, we arrive at the following effective 4D action:
\begin{equation}\label{spsymmeff2}
\begin{split}
 S_{eff}&=\int d^4 x \Bigl(\frac{1}{2}({\partial h_0})^2
+\sum_{n=1}^{\infty} \frac{1}{2}({\partial h_n})^2
-\sum_{n=1}^{\infty} \frac{1}{2}m^2_{n}h^2_n+
\frac{1}{2} M( h_0^2 H_0^2(0)+\\
&\quad +\sum_{n=1}^{\infty} 2h_0h_n H_0(0) H_n(0)) -\lambda(a_0
h^4_0+\sum_{n=1}^{\infty} 4a_n h^3_0 {h}_n)\Bigr)\; ,
\end{split}
\end{equation}
where $a_0$ and $a_n$ are again the overlap integrals
\eqref{a0c0}, \eqref{ancn}. In this case, the VEVs are
\begin{equation}\label{vev0}
v_0= \sqrt{\frac{Mk}{4a_0 \lambda}}\; ,
\end{equation}
\begin{equation}\label{vevn}
v_n= \frac{M {H}_0 (0){H}_n (0)v_0-4a_{n}\lambda v_0^3}{m^2_n}\; .
\end{equation}
We see from \eqref{vev0} that the mass $M$ must be small, $M\sim
\frac{v_{SM}^2}{k} \; .$ Using this estimate as well as
Eqs.~\eqref{scprofiles}---\eqref{scnormn}, we find, modulo a
factor of order one
\begin{equation}\label{estimate}
v_n\sim \frac{v_{SM}^3}{m^2_n}\left(\frac{m_n}{k}\right)^{\alpha
-1}\; .
\end{equation}
It is straightforward to show that the cancellation between the
leading order terms does not occur in Eq.~\eqref{vevn}, so the
estimate \eqref{estimate} is indeed valid. However, this value is
still very small from the viewpoint of physical applications
discussed in Section 3.2.

\section{Mass pattern of fermions}

\subsection{Quarks}
The action for free bulk fermions is
\begin{equation}\label{freedom}
S^{\Psi}_5=\int d^4 x dy\sqrt{g}
\left(ig^{MN}\bar{\Psi}\Gamma_{M}\nabla_N \Psi -m_{\Psi}
\bar{\Psi} \Psi \right)\; ,
\end{equation}
where $\Gamma_{M}$ are the 5D gamma matrices in adS space-time,
$\nabla_{M}$ is the covariant derivative, $m_{\Psi}$ is the
fermion bulk mass. One chooses the fermions transforming as $\Psi
(-y) =\pm \gamma_5 \Psi (y)$ under the orbifold $Z_2$ symmetry,
where the lower sign refers to $SU(2)_{L}$-doublets $Q$ and the
upper one to singlets $u$ and $d$. As a result, there are no left
zero modes of singlet quarks and right zero modes of doublets
\cite{Gherghetta, Grossman} and one arrives at the SM chiral
structure. Hereafter we consider zero modes of fermions. Their
profiles are given by \cite{Gherghetta, Grossman}
\begin{equation}\label{prferm}
Q_{0}(y)=N_{L}e^{(2-c_{L})ky}\; ,\quad
u_{0}(y)=N^{u}_{R}e^{(2-c^{u}_{R})ky}\; , \quad
d_{0}(y)=N^{d}_{R}e^{(2-c^{d})ky}\; .
\end{equation}
The constants $c_{L,R}$ are related to the fermion bulk masses,
$c_{R}=\frac{m_{\Psi}}{k}\;, ~c_{L}=-\frac{m_{\Psi}}{k} \; ,$ and
the normalization constants are
\begin{equation}\label{fermnorm}
N_{L}=\sqrt{\frac{(1-2c_{L})k}{e^{(1-2c_{L})k\pi R}-1}}\; , \quad
N^{u,d}_{R}=\sqrt{\frac{(1-2c^{u,d}_{R})k}{e^{(1-2c^{u,d}_{R})k\pi
R}-1}}\; .
\end{equation}
It is worth noting that with the account of the warp factor in
\eqref{freedom}, the effective profiles of the zero modes are
\begin{equation}\label{aralka}
\Psi_0 =N e^{(1/2-c)ky}\; .
\end{equation}
Hence, the zero modes are localized towards the IR and UV branes
for $c<1/2$ and $c>1/2$, respectively. Now, assuming that the
Higgs field lives in the bulk, we introduce its interaction with
fermions,
\begin{equation}\label{quarks5D}
\begin{split}
S^{q}_{5}=\int d^4xdy \sqrt{g}
\Bigl(\lambda_{ij}^{d}\bar{Q_i}Hd_j+\lambda_{ij}^{u}\bar{Q_i}\widetilde{H}u_j+h.c.\Bigr)\;
.
\end{split}
\end{equation}
Neglecting the excited KK modes of the Higgs field for the time
being and integrating Eq.~\eqref{quarks5D} over the extra
dimensional coordinate, we derive the effective 4D action:
\begin{equation}\label{quarkseff}
 S_{eff}^{q}=\int d^4x
\Bigl(\lambda_{ij}^{d}I^{d}_{0ij}\bar{d}_{Li}(x)d_{Rj}(x)h_{0}(x)+\lambda_{ij}^{u}I^{u}_{0ij}\bar{u}_{Li}(x)
u_{Rj}(x)h_{0}(x)+h.c.\Bigr)\; ,
\end{equation}
where $I^{u,d}_{0ij}$ are the overlap integrals of the zero modes
of the Higgs and quark fields,
\begin{equation}\label{overlap0}
I^{u,d}_{0ij}=\int^{\pi R}_{0} dy
\sqrt{g}H_0(y)Q_{0i}(y)d_{0j}(y)\; .
\end{equation}
Explicitly,
\begin{equation}\label{overlap0expl}
I^{u,d}_{0ij}=N_{Li}N^{u,d}_{Rj}N_0\frac{1-e^{(2-\alpha-c_{Li}-c^{u,d}_{Rj})k\pi
R}}{(\alpha+c_{Li}+c^{u,d}_{Rj}-2)k}\; .
\end{equation}
In the low-energy theory, the action \eqref{quarks5D} leads to the
quark mass matrix,
\begin{equation}\label{massquarks1}
M^{u,d}_{ij}={\lambda}_{ij}^{u,d}I_{0ij}^{u,d}v_{SM}\; .
\end{equation}
In our calculations we assume that the condition
$2-\alpha-c_{Li}-c_{Rj} <0$ is satisfied, so that the exponential
factor in Eq.~\eqref{overlap0expl} can be neglected. Furthermore,
we do not introduce the hierarchy between the 5D Yukawa couplings
and set $\lambda^{u,d}_{i,j} \sim k^{-1/2}$. Then, using
Eq.~\eqref{massquarks1}, we estimate the elements of the mass
matrix as follows,
\begin{equation}\label{massquarkssim}
M_{ij}^{u,d}\sim N_{Li} N_{Rj}^{u,d} \frac{v_{SM}}{k}\; .
\end{equation}
The hierarchy between quark masses and mixings is generated by the
hierarchy between the normalization constants $N_{Li}$, $N_{Rj}$,
which in turn is due to zero mode profiles. Aiming at
diagonalizing the mass term, we perform the unitary
transformations of left and right quarks (up- and down-quarks
independently) with corresponding matrices $A_{L}$ and $A_{R}$.
The latter satisfy the conditions
\begin{equation}\label{krolik}
A_{L}MM^{\dagger}A_{L}^{-1}=M_{diag}^2\; , \quad
A_{R}{M}^{\dagger}{M} {A}_{R}^{-1}=M_{diag}^2\; .
\end{equation}
These ensure that the quark mass matrix
\begin{equation}\label{sabirman}
m=A_{L}MA^{-1}_{R}
\end{equation}
is diagonal.

 With no hierarchy between the Yukawa couplings, the Hermitean matrix $MM^{\dagger}$
 has the following structure,
\begin{equation}\label{mm}
(MM^{\dagger})_{ij} \sim N_{Li}N_{Lj}\sum_{k=1}^{3}N^2_{Rk}
\frac{v^2_{SM}}{k^2}\; .
\end{equation}
So, only the normalization constants of the doublets $N_{Li}$ are
responsible for the hierarchy in Eq.~\eqref{mm}. We order them as
follows: $N_{L1}\ll N_{L2}\ll N_{L3}$. Then the elements of the
matrix $A_{L}$ are estimated by the the ratios of the constants
$N_{Li}$:
\begin{equation}\label{AL}
A_L \sim\left(
\begin{array}{cccc}
1 & \frac{N_{L1}}{N_{L2}}  & \frac{N_{L1}}{N_{L3}} \\
\frac{N_{L1}}{N_{L2}} & 1  & \frac{N_{L2}}{N_{L3}}\\
\frac{N_{L1}}{N_{L3}} & \frac{N_{L2}}{N_{L3}} & 1
\end{array}
\right)\; .
\end{equation}
By analogy, the Hermitean matrix $M^{\dagger}M$ is estimated as
\begin{equation}
M^{\dagger}M\sim
N_{Ri}N_{Rj}\sum_{k=1}^{3}N^{2}_{Lk}\frac{v^2_{SM}}{k^2}\; .
\end{equation}
Then, assuming the hierarchy of the normalization constants,
$N_{R1}\ll N_{R2}\ll N_{R3}$, we obtain
\begin{equation}\label{AR}
A_R \sim\left(
\begin{array}{cccc}
1 & \frac{N_{R1}}{N_{R2}}  & \frac{N_{R1}}{N_{R3}} \\
\frac{N_{R1}}{N_{R2}} & 1  & \frac{N_{R2}}{N_{R3}}\\
\frac{N_{R1}}{N_{R3}} & \frac{N_{R2}}{N_{R3}} & 1
\end{array}
\right)\; .
\end{equation}
Formally, this estimate is valid also for $N_{R1}\sim N_{R2}\sim
N_{R3}$, as is the case for down-quarks (see below). Finally,
using the estimates \eqref{massquarkssim}, \eqref{AL} and
\eqref{AR}, we estimate the quark masses:
\begin{equation}\label{sim}
m^{u,d}_{i}\sim N_{Li}N^{u,d}_{Ri}\frac{v_{SM}}{k}\; .
\end{equation}

Now, let us consider flavor mixing in the quark sector, which is
described by the Cabibbo--Kobayashi--Maskawa matrix. The latter is
given by
\begin{equation}\nonumber
C=A_{L}^d ({A^u_L})^{-1}\; .
\end{equation}
Since the matrices $A_{L}^d$ and $A^u_L$ have one and the same
general structure given by Eq.~\eqref{AL}, the CKM matrix is also
estimated by the right hand side of Eq.~\eqref{AL}. We now recall
the entries of the CKM matrix,
\begin{equation}\label{CKM}
|C|=\left(
\begin{array}{cccc}
0.97 & 0.23 & 0.0040\\
0.23 & 0.97 & 0.042\\
0.0081& 0.041& 0.99
\end{array}
\right)\; ,
\end{equation}
and compare them with Eq.~\eqref{AL}. We see that the right
pattern is obtained for
\begin{equation}\label{NL}
\frac{N_{L1}}{N_{L2}}\approx \frac{1}{10}\; ,\quad
\frac{N_{L2}}{N_{L3}}\approx \frac{1}{25}\; .
\end{equation}
\unitlength 1mm 
\linethickness{0.4pt}
\ifx\plotpoint\undefined\newsavebox{\plotpoint}\fi 
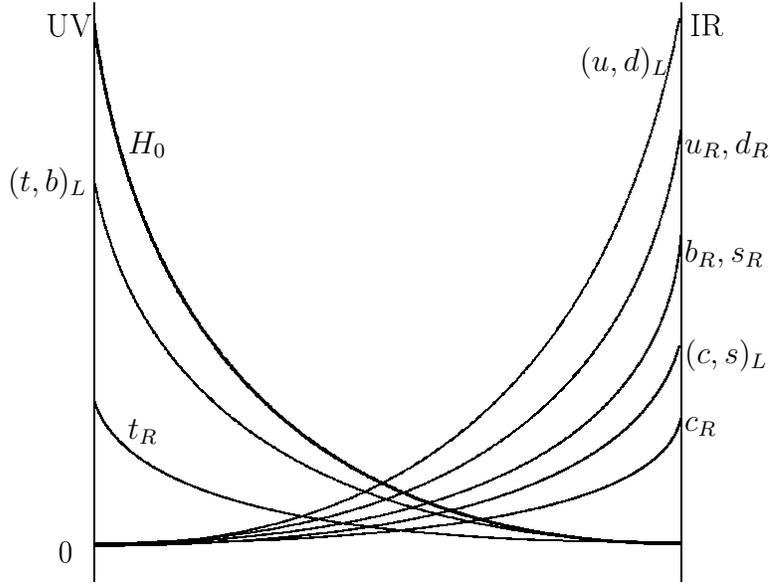
\begin{figure}
\begin{picture}(50,105)(15,40)
\put(62.5,153){\line(0,-1){77}} \put(140.5,153){\line(0,-1){77}}
\qbezier(62.5,81)(137.5,81.375)(140.5,122.25)
\qbezier(140.5,81.25)(68.5,80.875)(62.5,100)
\qbezier(62.5,81)(137,81.125)(140.5,97.75)
\qbezier(62.5,81)(132.5,81.375)(140.5,136.25)
\put(59.25,150){\makebox(0,0)[cc]{UV}}
\put(144,150){\makebox(0,0)[cc]{IR}}
\put(58.75,80){\makebox(0,0)[cc]{0}} \put(127,144){$(u,d)_{L}$}
\put(51,128){$(t,b)_{L}$} \put(67,95){$t_{R}$}
\put(141,105){$(c,s)_{L}$} \put(141,96){$c_{R}$}
\put(141,118.5){$b_{R},s_{R}$} \put(141,133){$u_{R},d_{R}$}
\put(67,133){$H_{0}$}
\qbezier(62.5,80.75)(132.5,81)(140.25,107.25)
\qbezier(62.5,80.75)(126.75,80.75)(140.25,150.75)
\qbezier(140.5,81)(71.5,82.375)(62.5,129.25)\thicklines
\qbezier(140.5,81)(72,81.250)(62.5,150)
\end{picture}
\vspace{-3.5cm} \caption{Effective profiles (see
Eq.~\eqref{aralka}) of quark zero modes.}\label{Klara}
\end{figure}
The estimate \eqref{sim} is consistent with known experimental
values of the quark masses \cite{pdg} $m_{u}\approx 2.6$ MeV,
$m_d\approx 5.0$ MeV, $m_c\approx 1.3$ GeV, $m_s\approx 100$ MeV,
$m_b\approx 4.2$ GeV, $m_t\approx 171$ GeV, provided that
\begin{equation}
\frac{N^{u}_{R1}}{N^{u}_{R2}}\approx \frac{1}{30}\; ,\quad
\frac{N^{u}_{R2}}{N^{u}_{R3}}\approx \frac{1}{5}\; ,\quad
\frac{N^{d}_{R1}}{N^{d}_{R2}}\approx \frac{1}{2}\; ,\quad
\frac{N^{d}_{R2}}{N^{d}_{R3}}\approx 1\; ,\quad
\frac{N^{d}_{R3}}{N^{u}_{R3}}\approx \frac{1}{40}\; .
\end{equation}
The overall scale of the normalization constants is obtained by
requiring that the mass of the top quark has the correct value.
This gives $N_{L3} N^{u}_{R3} \simeq k$. As an example, we choose
\begin{equation}
N^{u}_{L3}=1.3 \sqrt{k}\; , \quad N^{u}_{R3}=0.6 \sqrt{k}\; .
\end{equation}
Finally, using formulas \eqref{fermnorm}, one estimates the
dimensionless bulk masses of quarks, which we denoted by $c$.
Their values for the warp factor $k\pi R =10$ are given in
Table~\ref{muchacha}. We see explicitly that $c_{L,R}<1/2$ for the
lightest four quarks. As we noticed above, this means that they
live near the IR brane, as expected. Qualitative picture of the
quark localization is shown in Fig.~\ref{Klara}.

 We have still to choose the set of 5D Yukawa couplings $\lambda^{u,d}_{ij} \sqrt{k}$. Allowing them to vary within the interval
 $(1/3,~3)$, one can adjust
  masses and flavor mixings. Let us set
\begin{equation}\label{lambdau}
\lambda^{u}_{ij} \sqrt{k} = \left(
\begin{array}{cccc}
1.2 & 0.4  & -1.9 \\
1.7 & 1.1  & -0.9\\
-0.8 & 0.6 & 1.3
\end{array}
\right)\; ,
\end{equation}

\begin{equation}\label{lambdad}
\lambda^{d}_{ij} \sqrt{k} = \left(
\begin{array}{cccc}
0.9 & -0.4  & 1.3 \\
-1.3 & 1.3  & -0.5\\
1.8 & 0.3 & 1.1
\end{array}
\right)\; ,
\end{equation}
where we ignore phases for the sake of simplicity. With these 5D
Yukawa couplings one obtains the mixing matrix
\begin{equation}\label{CKM}
|C|= \left(
\begin{array}{cccc}
0.97 & 0.25 & 0.0040\\
0.25 & 0.97 & 0.040\\
0.0010& 0.040& 0.998
\end{array}
\right)
\end{equation}
and quark masses $m_{u}\approx 3$ MeV, $m_{d}\approx 9$ MeV,
$m_{c}\approx 1.5$ GeV, $m_{s}\approx 90$ MeV, $m_{b}\approx 4.5$
GeV, $m_{t}\approx 170$ GeV. Obviously, all these values are in a
reasonable agreement with the experimental data.
\begin{table}[!htb]
\hspace{6.5cm}
\begin{tabular}{|c|c|c|c|}
\hline
$c$ & $Q$ & $u$ & $d$\\
\hline
 $c_{L1}$ & -0.1 & - & -\\
\hline
 $c_{L2}$ & 0.2 & - & - \\
\hline
 $c_{L3}$ & 1.3 & - & -\\
\hline
 $c_{R1}$ & - & 0.0 & 0.0\\
\hline
 $c_{R2}$ & -& 0.4 & 0.1\\
\hline
 $c_{R3}$ & - & 0.7 & 0.1\\
\hline
\end{tabular}
\caption{ \label{muchacha} Quark parameters in the 5D SM with warp
factor $k\pi R=10$.}
\end{table}
\subsection{Leptons}

The interaction of leptons with the Higgs field in the 5D bulk is
\begin{equation}\label{leptons5D}
\begin{split}
S^{l}_5&=\int d^4xdy \sqrt{g}
\Bigl(\lambda^{l}_{ij}\bar{L_i}l_{j}+{\lambda}^{\nu}_{ij}\bar{L_i}\widetilde{H}
\nu_{j}+h.c.\Bigr)\; ,
\end{split}
\end{equation}
where $L_{i}(x,y)$ are the lepton $SU(2)_{L}$-doublets,
$l_{j}(x,y)$ and $\nu_j(x,y)$ are singlet charged and neutral
leptons, respectively, and $\lambda^{l}_{ij}$ and
$\lambda^{\nu}_{ij}$ are their 5D Yukawa couplings.

This interaction generates the neutrino masses of the Dirac type.
An interesting possibility here would be that neutrinos obtain
their masses predominantly via the interaction with excited KK
modes of the Higgs field. Then the smallness of the neutrino
masses would be due to the suppression of VEVs of the Higgs KK
excitations. Let us see, however, that this mechanism does not
work in the model we discuss. To this end, we keep all modes in
the decomposition of the Higgs field \eqref{scKK}. Inserting the
latter into Eq.~\eqref{leptons5D}, we derive the neutrino mass
matrix in the low-energy limit,
\begin{equation}\label{leptonsmasses}
M^{\nu}_{ij}= \lambda^{\nu}_{ij} \Bigl(v_0I^{\nu}_{0ij}
+\sum_{n=1}^{\infty} v_nI^{\nu}_{nij}\Bigr)\; ,
\end{equation}
where $v_0$ and $v_n$ are VEVs of the zero and excited modes of
the Higgs field, as described in Section 2; $I^{\nu}_{0ij}$ and
$I^{\nu}_{nij}$ are the overlap integrals of appropriate wave
functions. The first integral is given by the
Eq.~\eqref{overlap0expl} with the substitution $u \rightarrow
\nu$, while the second one is
\begin{equation}
I^{\nu}_{nij}=\int_{0}^{\pi R} dyL_{io}(y)H_{n}(y)\nu_{j}(y)\; ,
\end{equation}
where $H_{n}(y)$ is given by Eq.~\eqref{scprofn}. Explicitly,
\begin{equation}\label{lol}
I^{\nu}_{nij}=N_{n}N_{Li}N^{\nu}_{Rj}
\frac{1}{m_n}\left(\frac{m_n}{k}\right)^{c_{Li}+c^{\nu}_{Rj}-1}
\int^{\beta_n}_{\frac{m_n}{k}}  s^{1-c_{Li} -c^{\nu}_{Rj}}
J_{\alpha}(s)ds\; .
\end{equation}
Hereafter we assume that $c_{L}+c^{\nu}_{R} <2+\alpha$, so that
the integral here is of order $1$. One can show that our basic
conclusion is valid in the opposite case as well.

The expression \eqref{leptonsmasses} shows that neutrinos obtain
their masses due to their interactions with both zero mode and
excited modes of the Higgs field. The excited mode contribution
would dominate for
\begin{equation}\label{inequal1}
v_0 I^{\nu}_{0} \ll \sum_{n=1}^{\infty}v_nI^{\nu}_{n}\; .
\end{equation}
Omitting summation and using  \eqref{overlap0expl} and
\eqref{lol}, we rewrite Eq.~\eqref{inequal1} as follows,
\begin{equation}\label{inequal}
|e^{(2-\alpha -c_L -c^{\nu}_R)k\pi R} -1|\ll e^{(1-c_L
-c^{\nu}_R)k\pi R} \frac{v_n}{v_{0}}\; .
\end{equation}
Here we use the fact that $\frac{m_{n}}{k}e^{k\pi R}\sim 1$ for
the lightest KK modes. The condition \eqref{inequal} is equivalent
to the following two,
\begin{equation}
e^{(c_L +c^{\nu}_R -1)k\pi R}\ll \frac{v_n}{v_0}
\end{equation}
and
\begin{equation}\label{inequality}
e^{(1-\alpha)k\pi R}\ll \frac{v_n}{v_0}\; ,
\end{equation}
which must be satisfied simultaneously. We now recall the
expressions \eqref{vevnn} and \eqref{vevn} for VEVs of the excited
Higgs modes, and find that in both scenarios of spontaneous
symmetry breaking considered in Section 2, the inequality
\eqref{inequality} is not satisfied. In the best case, the
contribution of the Higgs KK excitations is suppressed by the
small factor $\frac{v^{2}_{SM}}{m^2_n}$ as compared to the zero
mode.
\unitlength 1mm 
\linethickness{0.4pt}
\ifx\plotpoint\undefined\newsavebox{\plotpoint}\fi 
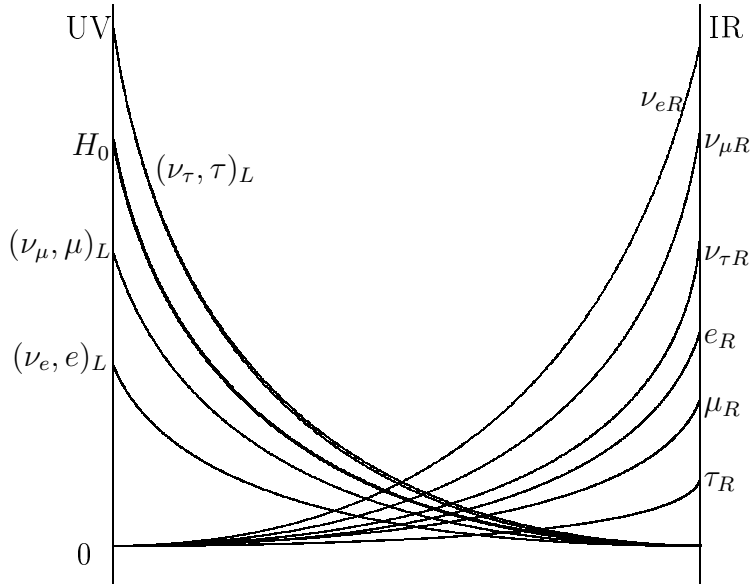
\begin{figure}
\begin{picture}(50,105)(15,40)
\put(62.5,153){\line(0,-1){77}} \put(140.5,153){\line(0,-1){77}}
\put(59.25,150){\makebox(0,0)[cc]{UV}}
\put(144,150){\makebox(0,0)[cc]{IR}}
\put(58.75,80){\makebox(0,0)[cc]{0}} \put(49,105){$(\nu_{e},
e)_{L}$} \put(68,130){$(\nu_{\tau}, \tau)_{L}$}
\put(57,133){$H_{0}$} \put(48.5,120){$(\nu_{\mu}, \mu)_{L}$}
\put(132.5,139.5){$\nu_{eR}$} \put(141,108){$e_{R}$}
\put(141,119){$\nu_{\tau R}$} \put(141,89){$\tau_{R}$}
\put(141,134){$\nu_{\mu R}$} \put(141,99){$\mu_{R}$}
\qbezier(140.5, 81)(72,81.250)(62.5,150)
\qbezier(62.5,81)(137.5,81.375)(140.5,122.25)
\qbezier(62.5,81)(132.5,81.375)(140.5,136.25)
\qbezier(62.5,81)(124.75,81.25)(140.5,147.5)
\qbezier(62.5,81)(131.5,80.75)(140.5,109.5)
\qbezier(62.5,81)(133.5,81.75)(140.5,100.5)
\qbezier(62.5,81)(137.5,81.125)(140.5,89.75)
\qbezier(140.5,81)(74.5,80.875)(62.5,120)
\qbezier(140.5,81)(72,81.125)(62.5,105)
\qbezier(140.5,81)(72,82.250)(62.5,150)\thicklines
\qbezier(140.5,81)(72,81.500)(62.5,135)
\end{picture}
\vspace{-3.5cm} \caption{Effective profiles (see
Eq.~\eqref{aralka}) of lepton zero modes.}\label{Klarushka}
\end{figure}

Thus, all lepton masses are obtained via the interaction with the
zero mode of the Higgs field. Still, the picture here is rather
different as compared to the quark sector. Indeed, the lepton
mixing matrix does not exhibit strong hierarchy \cite{pdg},
$$|C| \approx \left(
\begin{array}{cccc}
0.79-0.88 & 0.47-0.61 & <0.18\\
0.19-0.52 & 0.42-0.73 & 0.52-0.82\\
0.20-0.53 & 0.44-0.74 & 0.56-0.81
\end{array}
\right)\; .
$$
Similarly to the case of quarks, we estimate it by the right hand
side of Eq.~\eqref{AL}. Then we conclude that the normalization
constants $N_{L}$ are of one and the same order. It is therefore
natural to assume that all $N_{L}\sim \sqrt{k}$, so that the
dimensionless bulk masses of lepton doublets $c_L
>1/2$, i.e., the doublets reside near the UV brane. Otherwise, we
would need to fine tune the parameters $c_{L}$ to be very close to
each other.

Obviously, up to the change of notations $u\rightarrow \nu$ and
$d\rightarrow l$, the estimate Eq.~\eqref{sim} remains valid for
leptons. By choosing the normalization constants of singlet
fermions $N_{R}$ and thus the parameters $c^{l}_R$ and $c^{\nu}_R$
in an appropriate way, one adjusts 4D masses of leptons. Since we
assume that $N_{L}\sim \sqrt{k}$, the constants $N_{R}$ must be
small, $N_{R}\ll \sqrt{k}$, in order that the lepton masses be
small compared to the Higgs VEV. Hence, all $c^{l}_R$ and
$c^{\nu}_R$ must be smaller than $1/2$. So, we come to the
assignment that all singlet leptons reside towards the IR brane.
Finally, in Table~\ref{iducha} we present the set of the 5D
parameters $c$ leading to the correct hierarchy of lepton masses.
We choose the normal hierarchy of neutrino masses and assume no
degeneracy, i.e. $m_{1}\ll m_{2}$, $m_{2}\approx \sqrt{\Delta
m^{2}_{sol}}\approx 0.008$ eV and $m_{3}\approx \sqrt{\Delta
m^{2}_{atm}}\approx 0.05$ eV. The qualitative picture of lepton
localization in the 5D bulk is shown in Fig.~\ref{Klarushka}.
\begin{table}[!htb]
\hspace{5cm}
\begin{tabular}{|c|c|c|c|}
\hline
$c$ & $L$ & $\nu$ & $e$\\
\hline
 $c_{L1}$ & 1.0 & - & -\\
\hline
 $c_{L2}$ & 2.0 & - & -\\
\hline
 $c_{L3}$ & 3.0 & - & -\\
\hline
 $c_{R1}$ & -& $<-2.1$ & -0.8\\
\hline
 $c_{R2}$ & - & -2.1& -0.3\\
\hline
 $c_{R3}$ & - & -1.8 & 0.1\\
\hline
\end{tabular}
\caption{ \label{iducha} Lepton parameters in the 5D SM with warp
factor $k\pi R=10$.}
\end{table}

To conclude, masses and mixings in both quark and lepton sectors
are reproduced in the model we discuss without introducing large
or small parameters. The profiles of the fermion and Higgs zero
modes are naturally steep in the warped fifth dimension, which
translates into the strong hierarchies of masses and quark mixings
in the 4D world. The non-hierarchical pattern of neutrino mixings
is also natural with our choice of the localization of the left
lepton doublets.

\section{Kaon mixing}
\subsection{Kaon mixing mediated by excited Higgs field}
Unlike in the case of the Higgs field localized towards the IR
brane, the FCNC suppression is not at all automatic in models we
consider. The main source of FCNC in the model with the bulk Higgs
field, whose zero mode is localized near the UV brane, is the
interaction of light quarks with the KK excitations of the Higgs
field. This interaction is fairly strong because both zero modes
of light quarks and the Higgs KK modes are large near the IR
brane, so they overlap substantially.

Let us consider in detail the interaction of light down-quarks
with the KK excitations of the Higgs field. Since this interaction
is not diagonal in the flavor space, it leads to kaon mixing,
which is severely constrained by experiment. Integrating the
relevant terms in the action \eqref{quarks5D} over the fifth
coordinate, we arrive at the effective 4D action,
\begin{equation}\label{effex}
 S^{eff}_{quark}=\int d^4x
\Bigl(\sum_{n=1}^{\infty}
y^{d}_{nij}\bar{d}_{Li}(x)d_{Rj}(x)h_{n}(x)+h.c.\Bigr)\; .
\end{equation}
Here $y^{u,d}_{nij}$ are the effective 4D Yukawa couplings. They
are given by
\begin{equation}\label{Yukawa}
y_{nij}=\lambda^{d}_{ij}I^{d}_{nij}\; ,
\end{equation}
where $I^{d}_{nij}$ are the overlap integrals of the Higgs KK
excitations and zero modes of the singlet down-quarks and quark
doublets,
\begin{equation}\label{overlapn}
I^{d}_{nij}=\int dy \sqrt{g}Q_{0i}(y)H_n(y)d_{0j}(y)\; .
\end{equation}
Explicitly,
\begin{equation}\label{karas}
I^{d}_{nij}=N_{Li}N^{u,d}_{Rj}
N_{n}\frac{1}{m_n}\left(\frac{m_n}{k}\right)^{c_{Li}+c^{u,d}_{Rj}-1}
\int^{\beta_n}_{0}  s^{1-c_{Li} -c^{u,d}_{Rj}} J_{\alpha}(s)ds\; .
\end{equation}
As shown in Section 3.1, the lightest down-quarks reside towards
the IR brane. Thus, their normalization constants are
\begin{equation}
N_{Li}\approx \sqrt{(1-2c_{Li})k}e^{(c_{Li}-1/2)k\pi R}, \quad
N^{d}_{Rj}\approx \sqrt{(1-2c^{d}_{Rj})k}e^{(c^{d}_{Rj}-1/2)k\pi
R}\; .
\end{equation}
In this way we obtain the 4D Yukawa couplings of s- and d-quarks,
\begin{equation}\label{baku}
y_{nij}\approx
\lambda^{d}_{ij}\sqrt{\frac{(1-2c_{Li})(1-2c^{d}_{Rj})k}{\int^{\beta_n}_0sJ^2_{\alpha}(s)ds}}
\int^{\beta_n}_0
\Bigl(\frac{s}{\beta_n}\Bigr)^{1-c_{Li}-c^{d}_{Rj}}J_{\alpha}(s)
ds\; .
\end{equation}
Here the flavor indices are $i,j=1,2$, and the integrals are of
order 1. Hence, the Yukawa couplings are unsuppressed,
$y_{nij}\sim 1$. To obtain the Yukawa couplings of the physical
quark states, we perform the rotation of the quark fields with the
matrices $A_{L}$ and $A_{R}$. We obtain in the physical basis
\begin{equation}
\begin{split}
y'_{n12}& \approx a_{L11}(y_{n11}a^{-1}_{R12}+y_{n12}a^{-1}_{R22})\; ,\\
y'_{21}& \approx
a_{L22}(y_{n21}a^{-1}_{R11}+y_{n22}a^{-1}_{R21})\; ,
\end{split}
\end{equation}
where we neglect the $b$-quark contribution; the constants $a$ are
the entries of the matrices $A_{L}$ and $A_{R}$ estimated by
Eqs.~\eqref{AL},~\eqref{AR}. Obviously, the physical Yukawa
couplings are also unsuppressed, $y'_{n12} \sim y'_{n21} \sim 1$.
This precisely means that the RS-GIM mechanism does not work in
the case of the Higgs field residing near the UV brane.
Consequently, the only way to suppress dangerous FCNC is to assume
that the Higgs KK excitations have very large masses.
\unitlength 1mm 
\linethickness{0.4pt}
\ifx\plotpoint\undefined\newsavebox{\plotpoint}\fi 
\begin{figure}
\begin{picture}(50,85)(52,40)
\put(81.558,92.389){\line(1,0){.9797}}
\put(83.518,92.376){\line(1,0){.9797}}
\put(85.477,92.362){\line(1,0){.9797}}
\put(87.436,92.349){\line(1,0){.9797}}
\put(89.396,92.335){\line(1,0){.9797}}
\put(91.355,92.322){\line(1,0){.9797}}
\put(93.315,92.308){\line(1,0){.9797}}
\put(95.274,92.295){\line(1,0){.9797}}
\put(97.234,92.281){\line(1,0){.9797}}
\put(99.193,92.268){\line(1,0){.9797}}
\put(101.153,92.254){\line(1,0){.9797}}
\put(69,103.25){\vector(1,-1){.07}}\multiput(56.5,114.25)(.0382848392,-.0336906585){653}{\line(1,0){.0382848392}}
\put(70.80,80.5){\vector(-1,-1){.07}}\multiput(81.5,92.25)(-.0336990596,-.0368338558){638}{\line(0,-1){.0368338558}}
\put(115.50,103.375){\vector(-1,-1){.07}}\multiput(128,114.75)(-.03777777778,-.0337037037){675}{\line(-1,0){.03777777778}}
\put(115.15,79.625){\vector(1,-1){.07}}\multiput(102.25,92)(.03440054496,-.03371934605){734}{\line(1,0){.03440054496}}
\put(90,95){$h_{n}$} \put(61.5,112){$s_{L}$}
\put(61.5,77){$\overline{d}_{R}$}
\put(116,110){$\overline{s}_{R}$} \put(120,77){$d_{L}$}
\end{picture}
\begin{picture}(50,85)(15,40)
\put(72.68,92.43){\line(1,0){.9797}}
\put(74.639,92.416){\line(1,0){.9797}}
\put(76.599,92.403){\line(1,0){.9797}}
\put(78.558,92.389){\line(1,0){.9797}}
\put(80.518,92.376){\line(1,0){.9797}}
\put(82.477,92.362){\line(1,0){.9797}}
\put(84.436,92.349){\line(1,0){.9797}}
\put(86.396,92.335){\line(1,0){.9797}}
\put(88.355,92.322){\line(1,0){.9797}}
\put(90.315,92.308){\line(1,0){.9797}}
\put(92.274,92.295){\line(1,0){.9797}}
\put(60.5,103.25){\vector(1,-1){.07}}\multiput(48,114.25)(.0382848392,-.0336906585){653}{\line(1,0){.0382848392}}
\put(62.25,80.5){\vector(-1,-1){.07}}\multiput(73,92.25)(-.0336990596,-.0368338558){638}{\line(0,-1){.0368338558}}
\put(105.85,103.375){\vector(-1,-1){.07}}\multiput(118.5,114.75)(-.03777777778,-.0337037037){675}{\line(-1,0){.03777777778}}
\put(106,79.625){\vector(1,-1){.07}}\multiput(93.5,92)(.03440054496,-.03371934605){734}{\line(1,0){.03440054496}}
\put(81,95){$h_{n}$} \put(54,112){$s_{L}$}
\put(53,77){$\overline{d}_{R}$}
\put(109.5,112){$\overline{s}_{L}$} \put(111.5,77){$d_{R}$}
\put(-12,75){a)} \put(69,75){b)}
\end{picture}
\vspace{-3cm} \caption{Excited Higgs mediated processes leading to
kaon mixing.\label{aralk}}
\end{figure}
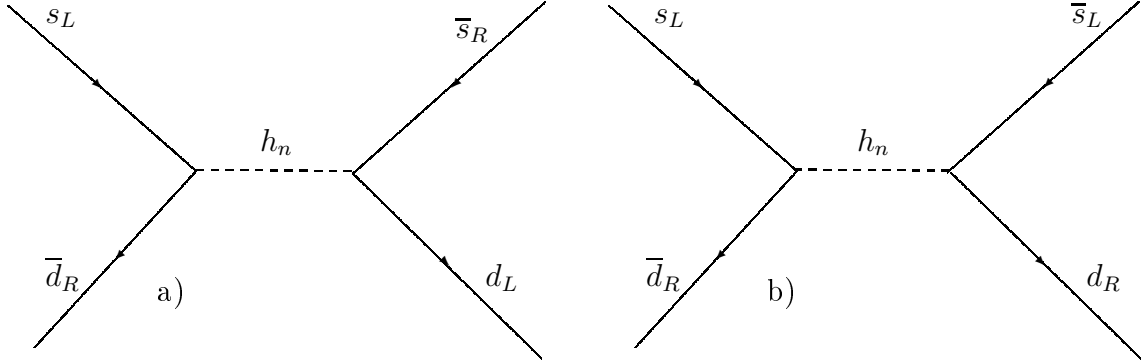

Generally, $\Delta F=2$ processes are described by the following
Hamiltonian:
\begin{equation}\label{Wilson}
H^{\Delta F=2}_{eff}=\sum_{a=1}^{5} C_a Q_a^{q_i q_j}+
\sum_{a=1}^{3} \widetilde{C}_a \widetilde{Q}_a^{q_i q_j}\; .
\end{equation}
The four-fermion operators $Q_{a}$ are given by
\begin{equation}\label{Wilsons}
Q_1^{q_i q_j}=\bar{q}^{\alpha}_{jL}\gamma_{\mu}q^{\alpha}_{iL}
\bar{q}^{\beta}_{jL}\gamma^{\mu}q^{\beta}_{iL}, \quad Q_2^{q_i
q_j}=\bar{q}^{\alpha}_{jR}q^{\alpha}_{iL}
\bar{q}^{\beta}_{jR}q^{\beta}_{iL}, \quad Q_3^{q_i
q_j}=\bar{q}^{\alpha}_{jR}q^{\beta}_{iL}
\bar{q}^{\beta}_{jL}q^{\alpha}_{iR}\; ,
\end{equation}
\begin{equation}\nonumber
Q_4^{q_i q_j}=\bar{q}^{\alpha}_{jR}q^{\alpha}_{iL}
\bar{q}^{\beta}_{jL}q^{\beta}_{iR}, \quad Q_5^{q_i
q_j}=\bar{q}^{\alpha}_{jR}q^{\beta}_{iL}
\bar{q}^{\beta}_{jL}q^{\alpha}_{iR}\; .
\end{equation}
The operators $\widetilde{Q}_a$ are obtained by the interchange
$L\leftrightarrow R$. Choosing $q_{j},q_{i} =d,s$, we focus on
kaon mixing. The exchange by the KK excitations of the Higgs field
contributes to the coefficients $C_2$, $\widetilde{C}_2$ and
$C_4$,
\begin{equation}\label{row24}
C_{2}=\sum^{\infty}_{n=1} \frac{({y'}_{n21}^{\star})^2}{m^2_n}\; ,
\quad \widetilde{C}_2=\sum^{\infty}_{n=1}
\frac{(y'_{n12})^2}{m^2_n}\; , \quad C_{4}=\sum^{\infty}_{n=1}
\frac{2{y'}_{n12}{{y'}^{\star}_{n21}}}{m^2_n}\; .
\end{equation}
This is shown in Fig.~\ref{aralk}, where Figs.~\ref{aralk}a and
~\ref{aralk}b correspond to $C_{2}$ and $C_{4}$, respectively. The
imaginary parts of the coefficients $C$ are responsible for
CP-violating mixing of kaons $K^{0}_1$ and $K^{0}_2$. Within our
model, there is no natural way to suppress the phases of the
coefficients $C$. Thus, we assume that $\mbox{Im}C\sim C$.

The experimental constraints on the imaginary parts are
\cite{Bona}
\begin{equation}\label{bon24}
-5.1 \times 10^{-17} ~\mbox{GeV}^{-2} \lesssim ~\mbox{Im} C_{2},
~\mbox{Im} \widetilde{C}_2 \lesssim 9.3 \times 10^{-17}
~\mbox{GeV}^{-2}\; ,
\end{equation}
\begin{equation}\label{bona24}
-1.8 \times 10^{-17} ~\mbox{GeV}^{-2}\lesssim ~\mbox{Im} C_{4}
\lesssim 0.9 \times 10^{-17} ~\mbox{GeV}^{-2}\; .
\end{equation}
By comparing Eq.~\eqref{bon24} with Eq.~\eqref{row24} and
recalling that $y'_{nij} \sim 1$, we see that the masses of the
Higgs KK excitations must be very large:
\begin{equation}\label{constraint!}
m_n\gtrsim  5\times 10^{5} ~\mbox{TeV}\; .
\end{equation}
The real parts of the coefficients $C$ contribute to the kaon mass
difference. The corresponding constraints are three orders of
magnitude weaker than \eqref{bon24} and \eqref{bona24}. Using
these constraints, we find that irrespectively of the above
assumption $\mbox{Im} C\sim C$, the masses of the Higgs KK
excitations must obey $m_{n} \gtrsim 10^{4}~\mbox{TeV}$.

\subsection{Other sources of kaon mixing}
A possible way to avoid the constraint \eqref{constraint!} is to
assume that the Higgs field is localized on the UV brane. Hence,
it does not have KK excitations at all, and the analysis of
Section 4.1 does not apply. In this case, the major source of kaon
mixing is the interaction of down-quarks with the KK excitations
of the bulk gauge fields. For simplicity, let us consider the
interaction with the bulk photons; exchange by the KK Z-bosons is
treated in a similar way and yields analogous results. The
relevant part of the 5D action is given by
\begin{equation}\label{actel}
S_{5}^{\gamma}=e_5\int d^4 x dy \sqrt{g}
\Bigl(\bar{Q}_{i}g^{MN}\Gamma_M A_N
Q_{i}+\bar{d}_{i}g^{MN}\Gamma_M A_N d_{i}\Bigr)\; .
\end{equation}
As usual, we expand the gauge field in the tower of KK modes,
\begin{equation}\label{klara}
A^{\mu}(x,y)=\sum_{n=0}^{\infty} a^{\mu}_{n}(x)A_{n}(y)\; .
\end{equation}
The zero mode of the bulk electromagnetic field is flat
\cite{Rizzo, Pomarol},
\begin{equation}\label{el0}
A_{0}(y)=\frac{1}{\sqrt{\pi R}}\; .
\end{equation}
Therefore, the 4D electric charge is $e=\frac{e_5}{\sqrt{\pi R}}$.
The profiles of the KK excitations are \cite{Rizzo, Pomarol}
\begin{equation}\label{eln}
A_{n}(y)=N^{\gamma}_{n}e^{ky}\left[J_{1}
\Bigl(\frac{m^{\gamma}_n}{k}e^{ky}\Bigr)
+C_{n}Y_{1}\Bigl(\frac{m^{\gamma}_n}{k}e^{ky}\Bigr)\right]\; ,
\end{equation}
where $m^{\gamma}_n$ denote the masses of the KK excitations. The
normalization factor $N^{\gamma}_n$ is given by
\begin{equation}\label{gaugenormn}
N^{\gamma}_{n}\approx \frac{m^{\gamma}_{n}}{\sqrt{k}}
\frac{1}{\sqrt{{\int}^{\gamma_n}_0 sJ^2_{1}(s)ds}}\; ,
\end{equation}
while the constant $C_n$ is
\begin{equation}
C_{n}=-\frac{J_{1}\Bigl(\frac{m^{\gamma}_n}{k}\Bigr)+\frac{m^{\gamma}_n}{k}J'_{1}\Bigl(\frac{m^{\gamma}_n}{k}\Bigr)}{Y_{1}\Bigl(\frac{m^{\gamma}_n}{k}\Bigr)+
\frac{m^{\gamma}_n}{k}Y'_{1}\Bigl(\frac{m^{\gamma}_{n}}{k}\Bigr)}\;
.
\end{equation}
The masses of the KK excitations of the electromagnetic field are
determined from the following eigenvalue equation \cite{Rizzo,
Pomarol}
\begin{equation}
\frac{m^{\gamma}_{n}e^{k\pi R}}{k} =\gamma_{n}, \quad
J_{1}\left(\frac{m^{\gamma}_{n}}{k}e^{k\pi R}\right)=0 \; .
\end{equation}

Hereafter we omit the zero mode in the decomposition
\eqref{klara}, since the interaction with the zero mode is
universal in the flavor space and does not give rise to flavor
violating processes. By inserting Eq.~\eqref{klara} into the 5D
action \eqref{actel} and integrating the latter over the fifth
coordinate, we arrive at the following effective 4D action:
\begin{equation}\label{effactel}
S_{eff}^{\gamma}=\int d^4 x \Bigl(\sum_{n=1}^{\infty}
b^{L}_{nij}\bar{d}_{Li}
(x)\gamma_{\mu}d_{Lj}(x)a^{\mu}_n(x)+(L\leftrightarrow R)\Bigr)\;
.
\end{equation}
The constants $b^{L(R)}_{nij}$ are the effective 4D couplings of
left (right) down-quarks with the $n$-th KK excitation. These
couplings are obtained from the initial 5D theory,
\begin{equation}
b^{L(R)}_{nij}=e_5 W^{L(R)}_{nij}\label{koshka}\; ,
\end{equation}
where the constants $W^{L(R)}_{nij}$ are the overlap integrals of
the appropriate wave functions,
\begin{equation}\label{klarik}
W^{L}_{nij}=\delta_{ij} \int dy
\sqrt{g}e^{ky}A_{n}(y)Q_{Li}(y)Q_{Lj}(y)
\end{equation}
and the analogous expression for the constants $W^{R}_{nij}$. By
performing the integration in Eq.~\eqref{klarik} and using
Eq.~\eqref{koshka}, we obtain
\begin{equation}
b^{L}_{nij}=e_5
\delta_{ij}N_{Li}N_{Lj}N_{n}\frac{1}{m^{\gamma}_n}\Bigl(\frac{m^{\gamma}_n}{k}\Bigr)^{c_{Li}+c_{Lj}-1}\int^{\gamma_n}_0s^{1-c_{Li}-c_{Lj}}J_1(s)ds\;
.
\end{equation}
The expression for the couplings $b^{R}_{nij}$ is obtained by the
replacing $N_{Li}\rightarrow N^{d}_{Ri}$ and $c_{Li}\rightarrow
c^{d}_{Rj}$. As in the previous case, all RS suppression factors
disappear for $i,j=1,2$. Thus, the couplings $b^{L(R)}_{n11}$ and
$b^{L(R)}_{n22}$ are of the order of the 4D electromagnetic
coupling $e$.

The effective action \eqref{effactel} is diagonal in the quark
fields. The flavor violating terms appear upon the rotation of the
quark fields by the matrices $A_{L}$ and $A_{R}$,
\begin{equation}
S^{eff}_{K-\widetilde{K}}=\int d^{4}x \Bigl( \sum_{n=1}^{\infty}
{b'}^{L}_{n12} \bar{d}_{L}(x)\gamma_{\mu}s_L(x)a^{\mu}_{n}(x)+
\sum_{n=1}^{\infty} {b'}^{L}_{n21}
\bar{s}_{L}(x)\gamma_{\mu}d_L(x)a^{\mu}_{n}(x)+(L\leftrightarrow
R)\Bigr)\; ,
\end{equation}
where, according to Eqs.~\eqref{AL},~\eqref{AR}, the couplings
${b'}^{L(R)}_{n12}$ and ${b'}^{L(R)}_{n21}$ are estimated as
\begin{equation}
 |{b'}^{L}_{n12}|\sim |{b'}^{L}_{n21}|\sim |b^{L}_{n11}-b^{L}_{n22}|\frac{N_{L1}}{N_{L2}}
\end{equation}
and
\begin{equation}
|{b'}^{R}_{n12}|\sim |{b'}^{R}_{n21}|\sim
|b^{R}_{n11}-b^{R}_{n22}|\frac{N^{d}_{R1}}{N^{d}_{R2}}\; .
\end{equation}
 Note that
the constants $b'^{L}_{n12}$ and $b'^{L}_{n21}$ are suppressed as
compared to the initial ones $b^{L}_{n11}$ and $b^{L}_{n22}$,
which is a consequence of the smallness of the factor
$\frac{N_{L1}}{N_{L2}}$. Note also that the profiles of $d_{R}$
and $s_{R}$ can be chosen very similar to each other, so that
$b^{R}_{n11}\approx b^{R}_{n22}$ and hence $b'_{n12}$ and
$b'_{n21}$ can be made very small. For the warp factor $k\pi R=10$
and the parameters $c$ listed in Table \ref{muchacha}, we obtain
$|{b'}^{L}_{n21}|\sim |{b'}^{L}_{n12}|\sim \frac{1}{50}$ and
$|{b'}^{R}_{n12}|\sim |{b'}^{R}_{n21}|\sim \frac{1}{30}$.

The exchange by the KK excitations of the electromagnetic field
gives contribution to the coefficients $C_{1}$ and
$\widetilde{C}_1$,
\begin{equation}\label{bon1}
 C_1=\sum_{n=1}^{\infty}
\frac{({b'}^{L}_{n12})^2}{{m^{\gamma}_n}^2}\; ,\quad
\widetilde{C}_1=\sum_{n=1}^{\infty}
\frac{({b'}^{R}_{n12})^2}{{m^{\gamma}_n}^2}\; .
\end{equation}
Their imaginary parts are constrained as follows \cite{Bona},
\begin{equation}
-4.4 \times 10^{-15} ~\mbox{GeV}^{-2}   \lesssim ~\mbox{Im} C_1,
 ~\mbox{Im} \widetilde{C}_{1}\lesssim 2.8 \times 10^{-15}
~\mbox{GeV}^{-2}\; .
\end{equation}
These bounds imply the following constraint on the masses of the
photon KK excitations:
\begin{equation}\label{constraint!!}
m^{\gamma}_{n}\gtrsim 700~\mbox{TeV}\; .
\end{equation}
We see that the masses of the KK excitations can be three orders
of magnitude smaller than the ones in the bulk Higgs scenario.
However, their values are still out of reach of future
experiments.

Besides the exchange by the KK modes, there are other sources of
FCNC. They are important in the case of the IR-localized Higgs
field, but subdominant in our scenario. One of these sources is
the interaction of down-quarks with the zero mode of the
$Z$-boson. The profile of the latter is not exactly flat
\cite{Rizzo, Pomarol}. However, the deviation from the flatness is
of the order of the ratio $\frac{m^{2}_{Z}}{m^{2}_{KK}}$, where
$m_{KK}$ is the typical mass scale of the KK excitations.
Accordingly, flavor-violating vertices are suppressed by
$\frac{m^{2}_{Z}}{m^{2}_{KK}}$.  With $m_{KK}$ constrained by
Eqs.~\eqref{constraint!} or~\eqref{constraint!!}, the contribution
to the coefficient $C_1$ coming from the interaction with the zero
mode of the Z-boson is negligibly small, $C_1\sim
\frac{m^2_{Z}}{M^4_{KK}}$.

As described in \cite{Azatov}, there are also flavor violating
processes mediated by the zero mode of the Higgs field. They occur
due to the interaction with the KK excitations of fermions. These
processes give negligibly small contribution to the coefficients
$C$ for the same reason as above.

\section{Acknowledgements}
The authors thank Valery Rubakov for useful discussions. This work
was supported by Federal Agency for Science and Innovation of
Russian Federation under state contracts 02.740.11.5194 (SM) and
02.740.11.0244 (MO and SR), by the Russian Foundation of Basic
Research grant 08-02-00287 (SM), by the grant NS-5525.2010.2 (MO
and SR), by the grants of the President of the Russian Federation
MK-4317.2009.2 (MO) and MK-7748.2010.2 (SR), by Federal Agency for Education under state
contract P520 (SR), by the Dynasty Foundation (SR).

\end{document}